\documentclass[aps,prb,reprint,superscriptaddress,floatfix]{revtex4-2}

\usepackage{amsmath}
\usepackage{amssymb}
\usepackage{xcolor}
\usepackage{bm}
\usepackage{graphicx}
\usepackage[percent]{overpic}

\usepackage[unicode=true,colorlinks=true]{hyperref}

\usepackage[export]{adjustbox}

\newcommand{\beq}{\begin{equation}}
\newcommand{\eeq}{\end{equation}}
\newcommand{\beqn}{\begin{align}}
\newcommand{\eeqn}{\end{align}}

\begin{document}

\title{Phase-Controlled Majorana Zero Modes in Altermagnetic Topological-Insulator Josephson Junctions}

\author{Hao Dong}
\affiliation{School of Physics and Institute for Quantum Science and Engineering, Huazhong University of Science and Technology, Wuhan, Hubei 430074, China}
\affiliation{Tsung-Dao Lee Institute and School of Physics and Astronomy, Shanghai Jiao Tong University, Shanghai 201210, China}

\author{Xun-Jiang Luo}
\affiliation{High Magnetic Field Laboratory, HFIPS, Chinese Academy of Sciences, Hefei, Anhui 230031, China}

\author{Xiao-Hong Pan}
\email{panxiaohong@jnu.edu.cn}
\affiliation{College of Physics and Optoelectronic Engineering, Department of Physics, Jinan University, Guangzhou 510632, China}

\author{Xin Liu}
\email{phyliuxin@sjtu.edu.cn}
\affiliation{Tsung-Dao Lee Institute and School of Physics and Astronomy, Shanghai Jiao Tong University, Shanghai 201210, China}
\affiliation{School of Physics and Institute for Quantum Science and Engineering, Huazhong University of Science and Technology, Wuhan, Hubei 430074, China}
\affiliation{Hefei National Laboratory, Hefei 230088, China}
\affiliation{Shanghai Research Center for Quantum Sciences, Shanghai 201315, China}

\begin{abstract}
We exploit facet-dependent Andreev phase shifts to control topological superconductivity with a phase bias in a three-dimensional altermagnetic topological-insulator Josephson junction. In the weak link between two conventional $s$-wave superconductors, the $d$-wave altermagnetic order produces anisotropic momentum shifts of the surface Dirac cones. The resulting net momentum of the states involved in Andreev reflection generates additional propagation phases that differ between facets. Consequently, the facet-resolved Andreev spectra exhibit gap closings at distinct phase biases, giving rise to topological superconducting regimes that host Majorana zero modes (MZMs). We further show that the spatial locations of the MZMs can be controlled by the phase bias. Moreover, these topological superconducting transitions are only weakly affected by moderate variations in the chemical potential, obviating the need for fine-tuning to the Dirac point. Our results establish a platform for realizing and spatially controlling MZMs by tuning the superconducting phase bias in altermagnetic topological-insulator Josephson junctions.
\end{abstract}

\maketitle

\section{Introduction}
\label{sec:intro}

Altermagnets are characterized by compensated magnetic order and momentum-dependent spin splitting despite a vanishing net magnetization~\cite{Noda2016,Hayami2019,Ahn2019,Yuan2020,Smejkal2020,Yuan2021,Libor2022,Libor2022a,Mazin2022,Krempasky2024}.
The combination of spin splitting and compensated magnetism has stimulated interest in their interplay with superconductivity, where the momentum dependence of the magnetic order can modify Cooper pairing and Josephson transport~\cite{zhang2024}.
In particular, recent studies have explored altermagnet-based superconducting heterostructures as platforms for realizing topological superconductivity and Majorana zero modes (MZMs)~\cite{Ghorashi2024,Hadjipaschalis2025,Li2023,Fu2026}.
The non-Abelian exchange statistics of MZMs make them potential building blocks for fault-tolerant topological quantum computation~\cite{Nayak2008,Alicea2012}, motivating efforts to control their spatial configurations as well as to establish their existence. For example, superconducting phase differences and magnetic-order orientations have been proposed as control parameters for the spatial
configuration of MZMs in superconducting heterostructures~\cite{Fu2008,Volpez2019,Pahomi2020,Zhou2020,Miao2025}.

\begin{figure}
  \centering
  \includegraphics[width=0.8\columnwidth]{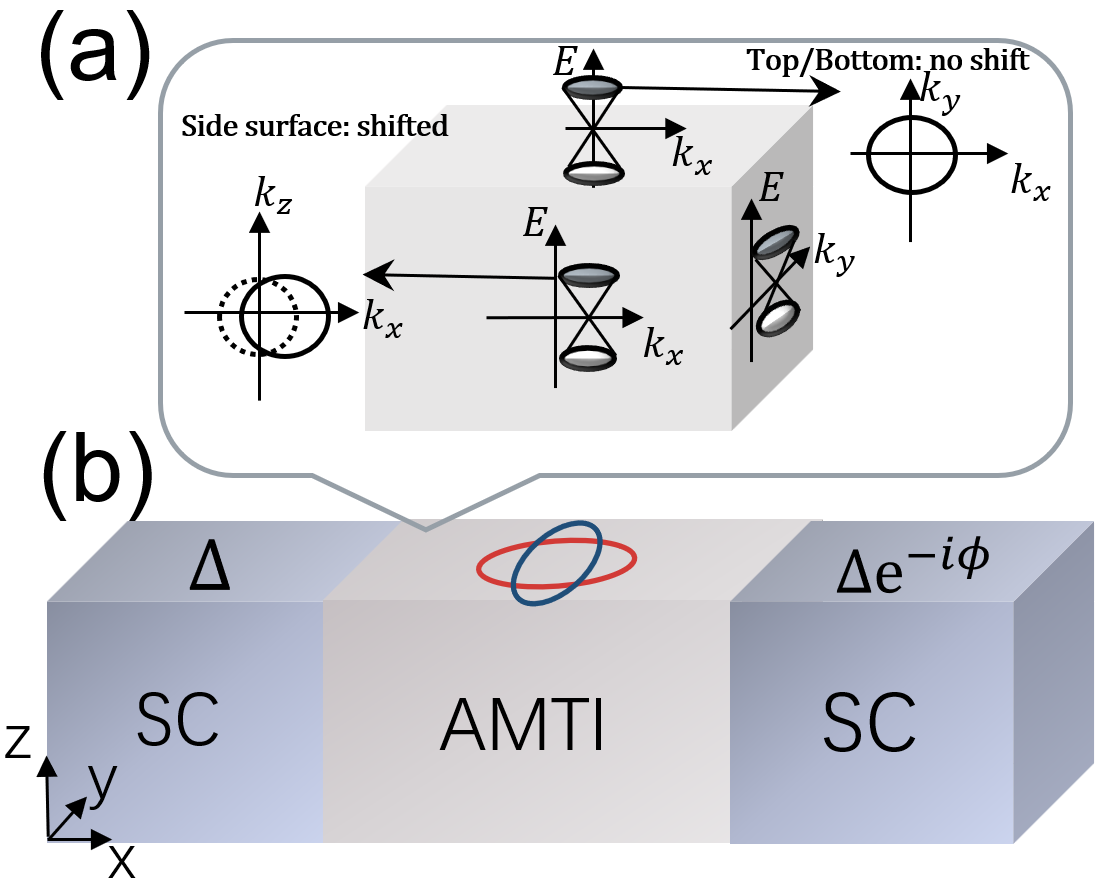}
\caption{\label{setup}
Schematic of the anisotropic AMTI surface states and device geometry.
(a) Facet-dependent Dirac-cone shifts induced by the altermagnetic order.
The Dirac cones on the lateral surfaces acquire finite momentum shifts, whereas those on the top and bottom surfaces remain unshifted to leading order.
(b) SC/3D AMTI/SC Josephson junction in which a three-dimensional AMTI forms the weak link between two conventional $s$-wave superconductors with phase difference $\phi$.
}
\end{figure}

Josephson junctions formed by conventional $s$-wave superconductors (SCs) and three-dimensional topological insulators (3D TIs) constitute an attractive platform for realizing phase-controlled topological superconductivity~\cite{Fu2008,Potter2013,Cook2012,Laubscher2024,schluck2024,Nikodem2025a}. 3D TIs possess a bulk band gap and metallic Dirac surface states with spin-momentum locking~\cite{Hasan2010,Zhang2009,Liu2010,Fu2008,Stanescu2010,Cook2011}. In a Josephson-junction geometry, tuning the superconducting phase difference modifies the surface Andreev spectrum and can drive topological phase transitions~\cite{Fu2008,Potter2013,Papaj2021,schluck2024}, and this topological phase can persist over a broad chemical-potential range~\cite{Cook2012,Papaj2021,schluck2024,Nikodem2025a}. This platform has motivated theoretical studies of Majorana-mediated Josephson effects, anomalous interference patterns, and Majorana modes bound to Josephson vortices~\cite{Potter2013,Laubscher2024}. Experimentally, phase-resolved tunneling spectroscopy has observed robust phase-driven gap closing and reopening over a broad range of chemical potentials~\cite{schluck2024}, while topological-insulator-nanowire Josephson junctions have realized surface-state-dominated nanoscale superconducting quantum interference device (nano-SQUID) behavior with flux-dependent interference and gate-tunable surface asymmetry~\cite{Nikodem2025a}. In an ideal symmetric TI Josephson junction, however, symmetry-related surface channels exhibit the same response to the superconducting phase difference. This raises the natural question of whether the anisotropy of altermagnetic order can provide facet selectivity and thereby enable phase-controlled topological superconductivity in Josephson junctions.

In this work, we demonstrate that the superconducting phase difference can drive an SC/AMTI/SC Josephson junction into a topological superconducting phase. Projection of the $d$-wave altermagnetic order onto the surface-state subspaces reveals a facet-dependent momentum shift: the Dirac cones on the lateral surfaces are displaced, whereas those on the top and bottom surfaces remain unshifted to leading order. On the lateral surfaces, the two opposite-spin states involved in Andreev reflection carry a finite net momentum that introduces an additional propagation phase. This additional phase differentiates the Andreev spectra of inequivalent facets and separates their topological transitions. Consequently, over a finite interval of the phase difference, adjacent facets acquire different effective superconducting masses, leading to MZMs at the hinges where these facets intersect. The topological transition remains weakly dependent on the chemical potential over a broad range within the bulk gap and therefore requires no fine-tuning to the Dirac point. We further show that varying the superconducting phase difference relocates the MZMs between different hinges, providing direct phase control over their spatial localization. These results establish the facet-dependent response of AMTI surface states as a mechanism for realizing and controlling topological superconductivity in three-dimensional AMTI Josephson junctions.

The remainder of this paper is organized as follows. Section~II introduces the AMTI model and derives the effective surface Hamiltonians on inequivalent facets. Section~III presents the SC/AMTI/SC Josephson junction and analyzes the facet-dependent Andreev spectra and the topological transitions driven by the superconducting phase difference. Section~IV establishes the emergence of MZMs at the hinges of a finite three-dimensional device and demonstrates their phase-controlled relocation. Finally, we conclude the paper in Sec.~V.

\section{Anisotropically Shifted Surface Dirac Cones}
We first investigate the surface-state properties of a 3D AMTI, focusing on the effect of the $d$-wave altermagnetic order. The Hamiltonian in momentum space take the form as ~\cite{Fu2026}
\begin{align}
H_{\rm N}(\bm k)={}&
\left[M_0+2\sum_n t_n(1-\cos k_n)\right]\sigma_z-\mu
\nonumber\\
&+\sum_n\lambda_n\sin k_n\,s_n\sigma_x
\nonumber\\
&+2t_J(\cos k_y-\cos k_x)s_z\sigma_z,
\end{align}
where the directional sums run over $n=x,y,z$. The operators $\sigma_i$ and $s_i$ are Pauli matrices acting in the orbital and spin spaces, respectively. The parameters $M_0$, $t_n$, $\lambda_n$, and $\mu$ denote the orbital mass, nearest-neighbor hopping amplitudes, spin-orbit-coupling strengths, and chemical potential, respectively, while $t_J$ is the strength of the altermagnetic order. We set the lattice constant and $\hbar$ to unity, assume isotropy in the $x$--$y$ plane, and take $t_x=t_y\equiv t_{\parallel}$ and $\lambda_x=\lambda_y$. For $t_J=0$, the model reduces to a time-reversal-symmetric 3D TI~\cite{Zhang2009,Liu2010}. We focus on the strong-TI regime driven by band inversion at the $\Gamma$ point, which hosts gapless helical Dirac cones on all surfaces~\cite{Zhang2009,Liu2010}.

\begin{figure}
  \centering
\includegraphics[width=1\linewidth]{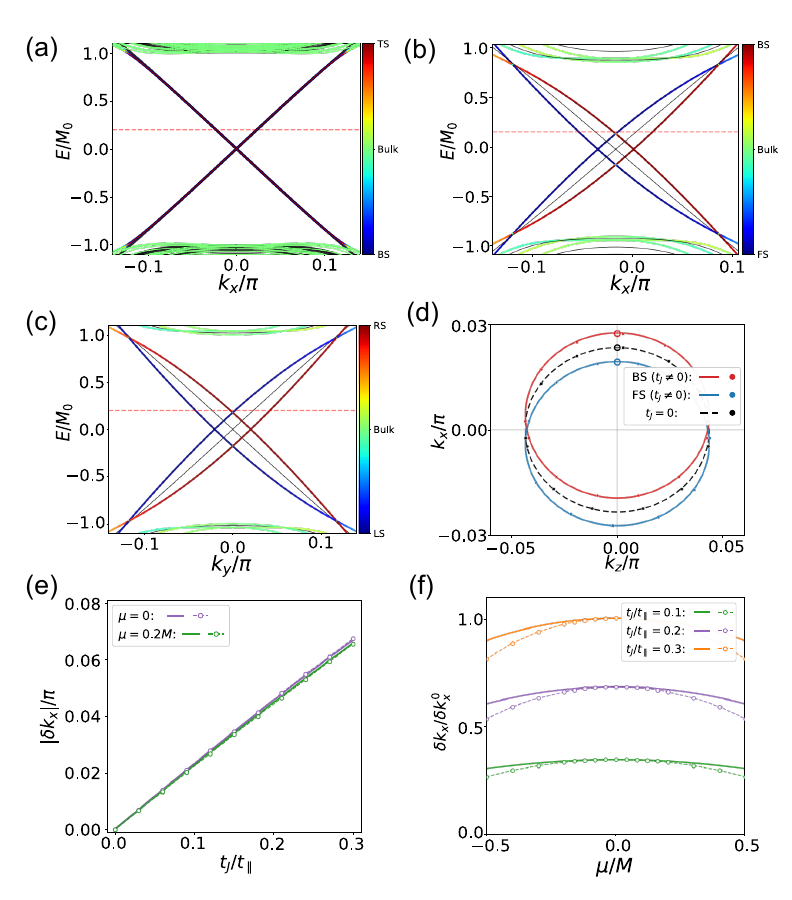}
\caption[Anisotropic surface spectra and effective momentum mismatch]{
Anisotropic surface spectra and the effective altermagnetic momentum mismatch. (a)--(c) Band structures of AMTI slabs with surfaces normal to the $z$, $y$, and $x$ directions, respectively. Black curves show the spectra for $t_J=0$, whereas the colored curves show those for finite $t_J$; the color scale encodes the localization of the states on the two opposite surfaces. The Dirac cones on the lateral surfaces are shifted, whereas the $z$-surface Dirac cone remains unshifted to leading order.
(d) Fermi contours on the two opposite $y$ surfaces: the red and blue curves correspond to the front surface (FS) and back surface (BS) at finite $t_J$, respectively, while the black dashed curve shows the contour for $t_J=0$. (e) Magnitude of the momentum mismatch $|\delta k_x|$ as a function of $t_J/t_{\parallel}$ at $\mu=0$ and $\mu=0.2M_0$.
The approximately linear increase illustrates that a finite chemical potential does not qualitatively modify the $t_J$ dependence. (f) Normalized momentum mismatch $\delta k_x/\delta k_x(\mu=0,t_J=0.3t_{\parallel})$ as a function of $\mu/M_0$ for several values of $t_J/t_{\parallel}$. Its weak dependence on $\mu/M_0$ demonstrates the robustness of the momentum mismatch against chemical-potential variations.
In panels (e) and (f), the colored curves correspond to the $t_J/t_{\parallel}$ values indicated in the legend. In panels (d)--(f), solid curves and dots denote results obtained from the analytical expressions and from direct numerical diagonalization of the full Hamiltonian, respectively.
The TI parameters are based on the four-band $k\cdot p$ model of the $\mathrm{Bi}_2\mathrm{Se}_3$ family~\cite{Zhang2009}, discretized with a lattice constant $a=1\,\mathrm{nm}$: $M_0=-0.15$ eV, $\lambda_x=\lambda_y=0.41$ eV, $\lambda_z=0.22$ eV, $t_x=t_y\equiv t_{\parallel}=0.566$ eV, and $t_z=0.1$ eV. Panels (a)--(d) use $t_J=10$ meV, while $t_J$ and $\mu$ are varied in panels (e) and (f) as indicated.
The parameter set is not intended as a quantitative fit to a specific material.}
  \label{surface}
\end{figure}

We examine the effect of the altermagnetic order on the surface-state dispersion by treating it as a perturbation and projecting it onto the surface-state subspace. Near the $\Gamma$ point, the altermagnetic term takes the form
\begin{align}
H_{\rm J}\simeq t_J(k_x^2-k_y^2)s_z\sigma_z.
\end{align}
For a surface normal to the $z$ direction, the projection of $H_{\rm J}$ onto the surface-state subspace vanishes identically~\cite{Fu2026}; further details are provided in Appendix. To verify this result numerically, we consider a slab
geometry with open boundary conditions (OBC) along $z$ and periodic boundary conditions (PBC) along $x$ and $y$, and calculate the band structure at $k_y=0$. Fig.~\ref{surface}(a) compares the spectra for $t_J=0$ (black curves) and finite $t_J$ (colored curves). The color scale encodes the localization of the eigenstates on the top and bottom $z$ surfaces. The surface bands for the two values of $t_J$ coincide exactly, in agreement with the analytical projection, confirming that the altermagnetic order leaves the $z$-surface states unchanged.

By contrast, the altermagnetic term has a finite projection onto the lateral surface-state subspace. For concreteness, we first consider a surface normal to the $y$ direction, described using a slab geometry with OBC along $y$ and PBC
along $x$ and $z$. Projecting onto the corresponding surface-state subspace gives the low-energy Hamiltonian
\begin{align}
H_{\rm eff}^{y} &= v_z k_z\widetilde{s}_x
-\left(v_{x} k_x-t_J k_x^2-\frac{t_JM_0}{t_{\parallel}}\right)\widetilde{s}_z-\mu.
\end{align}
Here, $\widetilde s_i$ ($i=x,y,z$) are Pauli matrices acting in the surface-state subspace, and $v_x\equiv\lambda_x$ and $v_z\equiv\lambda_z$. At $k_z=0$, the altermagnetic terms shift the surface-state dispersion along $k_x$, moving the Dirac point away from $k_x=0$. The numerical spectra in Fig.~\ref{surface}(b), plotted using the same conventions as in Fig.~\ref{surface}(a), confirm this result. At $t_J=0$, the two surface Dirac cones are degenerate and centered at $k_x=0$. Finite altermagnetic order lifts this degeneracy and shifts the two surface-state dispersions in opposite directions along $k_x$. Because the Hamiltonian possesses $\mathcal{T}C_{4z}$ symmetry---the combination of time-reversal and fourfold rotation symmetry---the altermagnetic order also shifts the surface Dirac cones on the $x$ facets, as shown in Fig.~\ref{surface}(c).

We next quantify the finite net momentum associated with the two opposite helical branches on the $y$ surfaces for chemical potentials within the bulk gap. The projected altermagnetic term has opposite signs on the front and back surfaces and therefore shifts the corresponding Fermi contours along $k_x$ in opposite directions, as illustrated by the red and blue contours in Fig.~\ref{surface}(d). On a given surface, the two opposite-spin Fermi points are consequently no longer related by $k_x\rightarrow-k_x$, resulting in a finite net momentum $\delta k_x\equiv k_{x,\uparrow}+k_{x,\downarrow}$. The sign of $\delta k_x$ is reversed on the opposite $y$ surface. At $k_z=0$, solving the surface-state dispersion for the two Fermi points and expanding around $\mu=0$ yields
\begin{align}
\delta k_x(\mu)
&\approx\delta k_x^{(0)}
\left[
1+\frac{M_0t_{\parallel}}{v_x^2}
\left(\frac{\mu}{M_0}\right)^2
\right],
\label{eq:deltak_exp}
\end{align}
where the net momentum at $\mu=0$ is
\begin{align}
\delta k_x^{(0)}
=\frac{v_x-\sqrt{v_x^2-4t_J^2M_0/t_{\parallel}}}{t_J}.
\end{align}
Expanding for weak altermagnetic order gives $\delta k_x^{(0)}\simeq 2M_0t_J/(v_xt_{\parallel})$, showing that $|\delta k_x^{(0)}|$ increases linearly with $t_J$ to leading order. As shown in Fig.~\ref{surface}(e), this approximately linear dependence on the altermagnetic strength persists at finite chemical potential. Furthermore, the leading chemical-potential correction in
Equation~\eqref{eq:deltak_exp} is quadratic in $\mu/M_0$. Accordingly, $\delta k_x$ changes only weakly as the chemical potential is varied away from the Dirac point, as shown more directly in
Fig.~\ref{surface}(f). The altermagnetism-induced finite net momentum is therefore a robust characteristic of the lateral surface states and does not require fine-tuning of the chemical potential.

\section{Engineering Anisotropic Andreev Bound States}

\begin{figure}
  \centering
  \includegraphics[width=\columnwidth]{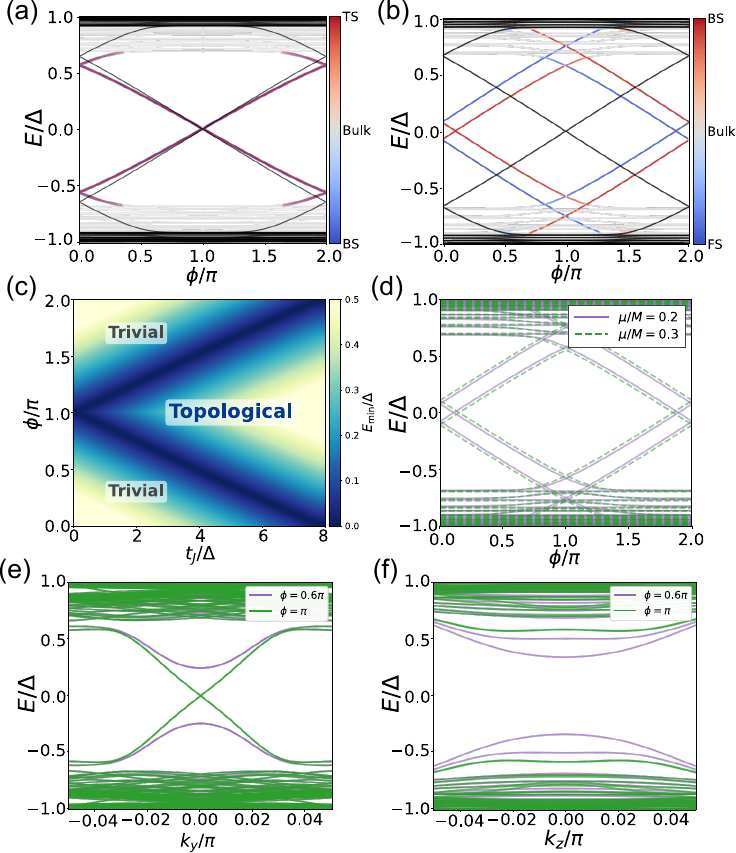}
\caption[Phase-driven topological transition in the SC/AMTI/SC junction]{
Phase-driven topological transition in the SC/AMTI/SC junction.
(a) and (b) Phase-dependent Andreev spectra of the surface Josephson channels on the $z$ and $y$ facets, denoted JJ-$z$ and JJ-$y$, at $k_y=0$ and $k_z=0$, respectively.
In both panels, black curves show the spectra for $t_J=0$, whereas colored curves show those for finite $t_J$; the colors encode the localization of the states on the two opposite facets.
The JJ-$z$ crossing remains at $\phi=\pi$, whereas the two JJ-$y$ crossings are shifted in opposite directions.
(c) Minimum positive excitation energy of JJ-$y$ as a function of $t_J$ and $\phi$.
The two shifted gap-closing lines delimit the interval in which the two $y$-facet Josephson masses have opposite signs.
(d) Phase-dependent JJ-$y$ spectra for several values of the chemical potential within the bulk gap, illustrating the weak chemical-potential dependence of the shifted crossings.
(e) and (f) Bogoliubov--de Gennes (BdG) dispersions of JJ-$z$ along $k_y$ and JJ-$y$ along $k_z$, respectively, at $\phi=0.6\pi$ and $\phi=\pi$.
At $\phi=0.6\pi$, both junction spectra are gapped over the displayed momentum ranges.
At $\phi=\pi$, JJ-$y$ remains gapped, whereas JJ-$z$ closes at $k_y=0$.
Unless otherwise specified, $\mu=0$.
The junction parameters are $L=65$ (total length along $x$), $\Delta=0.02$, $t_{sx}=t_{sy}=t_{sz}=0.1$, $t_J/\Delta=7$, $\mu_{\rm SC}/\Delta=4$, $L_x=25$, $L_y=40$, and $L_z=60$.
Energies are given in eV and lengths in units of $a=1\,\mathrm{nm}$.
The remaining TI parameters are as in Fig.~\ref{surface}.}
  \label{fig:phase}
\end{figure}

Recent experiments on lateral Josephson junctions based on three-dimensional topological insulators~\cite{schluck2024,Nikodem2025a} motivate us to consider a lateral SC/AMTI/SC junction in which two conventional $s$-wave superconductors are separated along the $x$ direction. As illustrated in Fig.~\ref{setup}(b), the AMTI normal region occupies $|x|<L_x/2$, and the superconducting order parameters on the two sides differ by a phase $\phi$. The exposed $y$ and $z$ facets extend across the normal region and therefore support surface Josephson channels. In the Nambu basis $\Psi=(C,is_yC^\dagger)^T$, where $C$ collects the electron operators of the heterostructure, the Bogoliubov--de Gennes (BdG) Hamiltonian is
\begin{equation}
\label{eq:BdG_Full}
H_{\rm BdG}(\phi)=
\begin{pmatrix}
H_{\rm e} & \Delta(x)\\[4pt]
\Delta^\dagger(x) & -s_yH_{\rm e}^{*}s_y
\end{pmatrix}.
\end{equation}
Here, $H_{\rm e}=H_{\rm N}+H_{\rm SC}+H_c$ is the normal-state
electron Hamiltonian. The term $H_{\rm N}$ is the real-space form of the AMTI Hamiltonian introduced in Sec.~II, $H_{\rm SC}$ describes the normal-state part of the superconducting electrodes, and $H_c$ denotes the coupling across the SC/AMTI interfaces. The normal-state Hamiltonian of the superconductors is 
\begin{align}
H_{\rm SC}={}&\sum_{\bm r\in{\rm SC}}
d_{\bm r}^{\dagger}
\left(2\sum_{n=x,y,z}t_{sn}-\mu_{\rm SC}\right)d_{\bm r}
\nonumber\\
&-\sum_{\bm r\in{\rm SC}}\sum_{n=x,y,z}
\left(
d_{\bm r+\hat{\bm n}}^{\dagger}
t_{sn}d_{\bm r}
+{\rm H.c.}
\right).
\end{align}
Here $d_{\bm r}$ is a four-component spin-orbital annihilation operator, $t_{sn}$ is the nearest-neighbor hopping amplitude along the $n$ direction, $\mu_{\rm SC}$ is the chemical potential of the superconducting electrodes, and $\hat{\bm n}$ is a unit lattice vector along $n$. The interface coupling is modeled by
\begin{equation}
H_c=-t_c\sum_{\langle\bm r,\bm r'\rangle_{\rm int}}
\left(
c_{\bm r}^{\dagger}d_{\bm r'}+{\rm H.c.}
\right),
\end{equation}
where $c_{\bm r}$ is the four-component spin-orbital annihilation operator in the AMTI and the sum runs over nearest-neighbor AMTI--SC sites across the interfaces at $x=\pm L_x/2$. The coupling $t_c$ preserves both spin and orbital indices. The superconducting pair potential is chosen as
\begin{equation}
\Delta(x)=
\begin{cases}
\Delta, & x>L_x/2,\\
0, & |x|<L_x/2,\\
\Delta e^{-i\phi}, & x<-L_x/2,
\end{cases}
\label{eq:pairing_profile}
\end{equation}
When the chemical potential lies within the AMTI bulk gap, the
low-energy states predominantly reside on the topological surface states. The surface states at the two $x$-normal SC/AMTI interfaces acquire a superconducting gap through the proximity effect. Along the facets normal to the $y$ and $z$ directions, the surface states extend across the AMTI region between the two superconductors, thereby forming surface Josephson junctions. The junctions on the top and bottom facets ($\pm z$) are collectively denoted JJ-$z$, whereas those on the front and back facets ($\pm y$) are denoted JJ-$y$.

The JJ-$z$ geometry has open boundaries along $x$ and $z$ and remains translationally invariant along $y$, with $k_y$ as a good quantum number. Conversely, the JJ-$y$ geometry has open boundaries along $x$ and $y$ and remains translationally invariant along $z$, with $k_z$ as a good quantum number. We first set the chemical potential at the Dirac point, $\mu=0$, and consider finite chemical potentials below. We denote the conserved transverse momentum by $k_\perp$, with $k_\perp=k_y$ for JJ-$z$ and $k_\perp=k_z$ for JJ-$y$. Particle-hole symmetry maps $E(k_\perp)$ to $-E(-k_\perp)$, making $k_\perp=0$ a particle-hole-invariant sector. We therefore first analyze the phase-dependent gap closings at $k_\perp=0$.

We first consider the reference limit $t_J=0$. Both JJ-$z$ and JJ-$y$ then reduce to the surface Josephson junctions of a time-reversal-symmetric three-dimensional topological insulator~\cite{Fu2008,Fu2009a}. In the short-junction limit, the low-energy spectrum of each surface channel
can be written as
\begin{equation}
E_{\pm}^{(0)}(k_{\perp},\phi)
\simeq
\pm\sqrt{
v_{\perp}^{\,2}k_{\perp}^{2}
+m_0^{2}(\phi)
}.
\end{equation}
Here $v_\perp=v_y\equiv\lambda_y$ for JJ-$z$ and $v_\perp=v_z\equiv\lambda_z$ for JJ-$y$. The phase-dependent Josephson mass is $m_0(\phi)=\Delta_{\rm eff}\cos(\phi/2)$, where $\Delta_{\rm eff}$ denotes the proximity-induced gap of the surface states. At zero transverse momentum, continuously following the signed Andreev branches gives
\begin{equation}
E_{\pm}^{(0)}(\phi)
\simeq
\pm\Delta_{\rm eff}
\cos\left(\frac{\phi}{2}\right).
\end{equation}
As $\phi$ is tuned through $\pi$, the phase-dependent mass $m_0(\phi)$ passes through zero and changes sign, corresponding to a mass inversion of the surface Josephson channel. At $\phi=\pi$, the superconducting phase configuration is invariant under time reversal, $\mathcal{T}=is_y\mathcal{K}$, where $\mathcal{K}$ denotes complex conjugation. Consequently, each surface Josephson channel hosts a pair of counterpropagating gapless Majorana modes~\cite{Fu2008,Fu2009a,liu2011}. In the absence of altermagnetic order, the two opposite-surface channels within each junction are degenerate, and their zero-energy crossings coincide at $\phi=\pi$. The corresponding spectra of JJ-$z$ and JJ-$y$ are shown by the black curves in Figs.~\ref{fig:phase}(a) and~\ref{fig:phase}(b), respectively, confirming this low-energy picture. In these numerical calculations, we employ an enlarged parent-superconductor gap $\Delta$ to shorten the coherence length, treating it as a model energy scale rather than a quantitative value for a particular conventional superconductor. Nevertheless, for the interface coupling considered here, the covered AMTI surfaces lie in the strong-proximity regime, yielding an induced gap $\Delta_{\rm ind}$ close to $\Delta$ and thus justifying $\Delta_{\rm eff}\simeq\Delta$. We now consider the effect of the altermagnetic order on the Josephson junction. As shown in Sec.~II, the $z$-surface states remain essentially unaffected by the altermagnetic order. The low-energy Andreev spectrum of JJ-$z$ is therefore nearly unchanged, and its zero-energy crossing remains pinned at $\phi=\pi$, as indicated by the colored curves in Fig.~\ref{fig:phase}(a). The response of JJ-$y$ is qualitatively different. On each $y$ surface, the two opposite-spin states participating in Andreev reflection carry a finite net momentum $\delta k_x$, whose sign is reversed between the front and back surfaces, as illustrated by the red and blue Fermi contours in Fig.~\ref{surface}(d). Cooper-pair propagation across the AMTI region consequently acquires an additional surface-dependent phase $\nu\delta\phi_{\rm AM}$, where $\delta\phi_{\rm AM}=\delta k_xL_x$ and $\nu=\pm1$ labels the front and back surfaces. In the short-junction limit, this effect is directly captured by the low-energy Andreev levels
\begin{equation}
E_{\nu,\pm}^{y}(\phi)
\simeq
\pm\Delta_{\rm eff}
\cos\left(
\frac{\phi+\nu\delta\phi_{\rm AM}}{2}
\right).
\end{equation}
The opposite phase shifts lift the surface degeneracy at $\phi=\pi$ and move the zero-energy crossings to $\phi_{c,\nu}=\pi-\nu\delta\phi_{\rm AM}$, as shown by the colored curves in Fig.~\ref{fig:phase}(b). The altermagnetic phase shift therefore separates the two JJ-$y$ gap closings, whereas the two JJ-$z$ channels remain degenerate and close simultaneously at $\phi=\pi$.

With $m_{y,\nu}(\phi)=\Delta_{\rm eff}\cos[(\phi+\nu\delta\phi_{\rm AM})/2]$ and $m_z(\phi)=m_0(\phi)$, the interval between the shifted JJ-$y$ crossings is characterized by $m_{y,+}m_{y,-}<0$. To compare the relative mass signs on different facets, we choose the local surface-state bases such that the four surface Josephson channels are expressed in a common Dirac convention along the perimeter of the $y$--$z$ cross section. This convention is fixed by requiring all four facet masses to reduce to the same $m_0(\phi)$ in the reference limit $t_J=0$. A change in the mass sign between adjacent facets then defines a Dirac-mass domain wall at their common hinge~\cite{Fu2008}. For $\phi\neq\pi$ within the interval between the shifted crossings, $m_z(\phi)\neq0$. Because the two $z$ facets carry the same mass $m_z(\phi)$, exactly one of the two $y$ facets has a mass of the opposite sign. The two hinges adjoining this $y$ facet are consequently mass domain walls and host MZMs. At $\phi=\pi$, the $z$-facet mass vanishes; as $\phi$ passes through this surface critical point, $m_z(\phi)$ reverses sign and the mass mismatch switches to the opposite $y$ facet. Fig.~\ref{fig:phase}(c) locates the two JJ-$y$ gap-closing lines through the minimum positive BdG eigenenergy as functions of $\phi$ and $t_J$. These lines delimit the interval $m_{y,+}m_{y,-}<0$, and their separation increases with $t_J$. As shown in Sec.~II, $\delta k_x$ depends only weakly on the chemical potential. Consistently, Fig.~\ref{fig:phase}(d) shows that the shifted crossing phases remain nearly unchanged over the range of $\mu$ considered. A fully gapped surface spectrum also requires the JJ-$z$ and finite-transverse-momentum sectors to remain gapped. We therefore calculate the BdG dispersions of JJ-$z$ along $k_y$ and JJ-$y$ along $k_z$, as shown in Figs.~\ref{fig:phase}(e) and~\ref{fig:phase}(f), respectively. At $\phi=0.6\pi$, both JJ-$z$ and JJ-$y$ remain gapped over the displayed momentum ranges. At $\phi=\pi$, JJ-$y$ remains gapped, whereas JJ-$z$ closes at $k_y=0$, consistent with $m_z(\pi)=0$. Thus, $\phi=0.6\pi$ represents a fully gapped surface-mass configuration, whereas $\phi=\pi$ is a JJ-$z$ surface critical point separating the two gapped configurations on its two sides.

\begin{figure}
  \centering
  \includegraphics[width=\columnwidth]{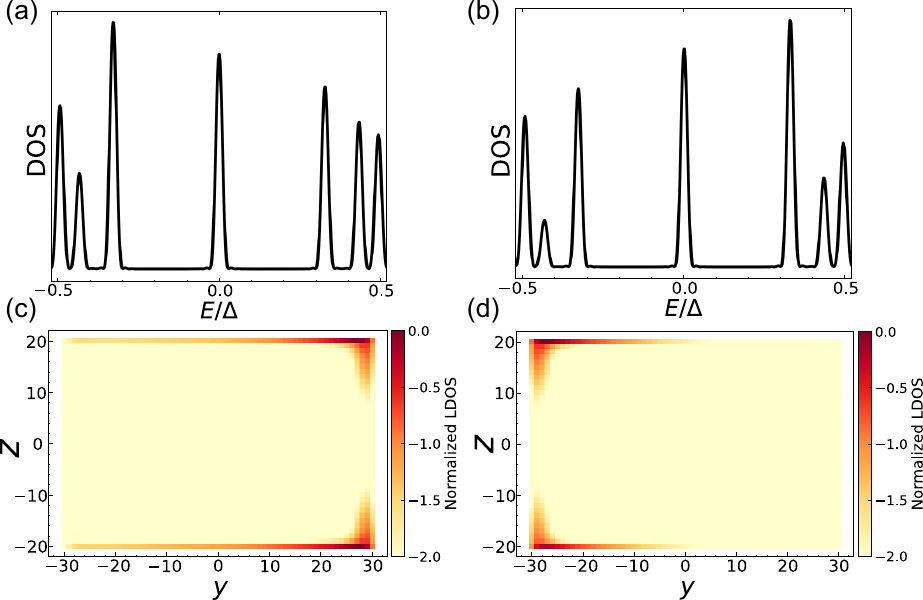}
\caption[Spatial distribution and phase control of Majorana zero modes]{
Spatial distribution and phase control of Majorana zero modes.
(a) and (b) DOS at superconducting phase differences $\phi=0.6\pi$ and $\phi=1.4\pi$, respectively.
Both spectra exhibit a pronounced zero-energy peak.
(c) and (d) Zero-energy LDOS projected onto the $y$--$z$ cross section by summing over the $x$ direction at $\phi=0.6\pi$ and $\phi=1.4\pi$, respectively.
The color scale represents $\log_{10}[\rho(E=0,\bm r)/\rho_{\max}]$, where $\rho_{\max}=\max_{\bm r}\rho(E=0,\bm r)$.
The zero-energy spectral weight is localized at the hinge mass domain walls, consistent with the Majorana zero modes predicted by the surface-mass analysis.
The hinge pair hosting the zero-energy modes switches as the phase is tuned across $\phi=\pi$.
Parameters are as in Fig.~\ref{fig:phase}.}
  \label{fig:ldos}
\end{figure}

\section{Majorana Zero Modes}
\label{sec:hinge}

We now turn to a finite three-dimensional device with open boundary conditions along all three spatial directions. To examine the boundary states associated with the two gapped surface-mass configurations identified above, we calculate the density of states (DOS) and the spatially resolved zero-energy local density of states (LDOS) at $\phi=0.6\pi$ and $\phi=1.4\pi$, which lie on opposite sides of the JJ-$z$ surface gap closing at $\phi=\pi$. The DOS shown in Figs.~\ref{fig:ldos}(a) and~\ref{fig:ldos}(b) exhibits a pronounced zero-energy peak at each phase difference, indicating the emergence of zero-energy bound states. Their spatial distributions are resolved by the zero-energy LDOS shown in Figs.~\ref{fig:ldos}(c) and~\ref{fig:ldos}(d): the spectral weight is concentrated at two hinges that correspond to the mass domain walls identified in Sec.~III, supporting the identification of the bound states as hinge Majorana zero modes. The LDOS distributions at the two phase differences show that the zero-energy spectral weight switches from one hinge pair to the other across $\phi=\pi$. Such a phase-dependent redistribution is expected to produce a corresponding modulation of the local zero-bias tunneling conductance measured at a fixed hinge~\cite{Ikegaya2021,Li2026}.

\section{Conclusion}
\label{sec:conclusion}
In this work, we have proposed a phase-controlled route to topological superconductivity in an SC/AMTI/SC Josephson junction. The anisotropic response of the topological surface states to the $d$-wave altermagnetic order drives topological superconducting transitions and gives rise to MZMs at the hinges. The MZMs can be created and spatially manipulated by the superconducting phase difference without requiring fine-tuning of the chemical potential to the Dirac point. The phase-controlled relocation of the MZMs also provides a direct experimental signature. A scanning tunneling microscope (STM) tip fixed at a hinge could measure the local differential conductance $dI/dV$. As $\phi$ is varied, the zero-bias conductance peak is expected to follow the MZMs from one hinge to another, yielding a phase-correlated switching pattern that is difficult for a static trivial Andreev bound state to reproduce~\cite{Volpez2019,Zhang2021TRI,Ikegaya2021,Miao2025,Li2026}.

Our proposal could be realized either in an intrinsic AMTI that combines an inverted bulk gap with altermagnetic order or in an engineered heterostructure. In the latter route, a conventional three-dimensional topological insulator, such as Bi$_2$Se$_3$ or Bi$_2$Te$_3$~\cite{Zhang2009}, could be combined with an altermagnetic layer and conventional $s$-wave superconducting contacts. While the altermagnetic order considered in our model is of $d$-wave type, the facet-selective mechanism may extend to other altermagnetic symmetries. A $g$-wave altermagnetic term may have different projections in the surface-state subspaces of inequivalent facets, thereby potentially generating facet-dependent momentum shifts similar to those studied here. MnTe and CrSb, experimentally reported as $g$-wave altermagnets, are also potential candidates for the altermagnetic component of an engineered heterostructure~\cite{Krempasky2024,Ding2024CrSb}. The recently proposed altermagnetic proximity effect supports the transfer of momentum-dependent spin splitting into an adjacent nonmagnetic layer~\cite{zhu2026}. Layered Rb$_{1-\delta}$V$_2$Te$_2$O and
epitaxial Mn$_5$Si$_3$ are possible altermagnetic components for exploring this route~\cite{Zhang2025,Reichlova2024Mn5Si3}. Introducing controlled inversion asymmetry through asymmetric interfaces or electrostatic gating may additionally enable a superconducting diode response in the present device geometry~\cite{Souto2022,Lu2023,Cheng2024,Boruah2025,Banerjee2024,Nikodem2025}, although this possibility lies beyond the scope of the present work.

\begin{acknowledgments}
We acknowledge useful discussions with Guo-Liang Guo and Li Chen. X. Liu acknowledges support from the Innovation Program for Quantum Science and Technology of China (Grant No. 2021ZD0302700), the National Natural Science Foundation of China (NSFC; Grant Nos. 92565305 and 12074133), the Shanghai Science and Technology Innovation Action Plan (Grant No. 24LZ1400800), and the Cultivation Project of the Shanghai Research Center for Quantum Sciences (Grant No. LZPY2024). X.-H. Pan acknowledges support from the NSFC (Grant No. 12504190) and the Guangdong Basic and Applied Basic Research Foundation (Grant No. 2026A1515011510).
\end{acknowledgments}

\appendix
\section{Low-Energy Effective Hamiltonian of the Altermagnetic Topological Insulator}

\subsection{Continuum Model}

We derive the low-energy surface Hamiltonians by expanding the lattice
model introduced in Sec.~II around the $\Gamma$ point. In the basis
\begin{equation}
\Phi_{\bm k}=
(c_{1\bm k\uparrow},c_{2\bm k\uparrow},
c_{1\bm k\downarrow},c_{2\bm k\downarrow})^{T},
\end{equation}
where $1$ and $2$ denote two low-energy orbitals with opposite inversion
parity, the continuum Hamiltonian reads
\begin{align}
H_{\rm N}(\bm k)=&M(\bm k)s_0\sigma_z-\mu s_0\sigma_0
+\lambda_xk_xs_x\sigma_x+\lambda_yk_ys_y\sigma_x
\nonumber\\
&+\lambda_zk_zs_z\sigma_x
+t_J(k_x^2-k_y^2)s_z\sigma_z,
\label{eq:app_continuum_amti}
\end{align}
with
\begin{equation}
M(\bm k)=M_0+t_xk_x^2+t_yk_y^2+t_zk_z^2.
\end{equation}
Here $s_i$ and $\sigma_i$ are Pauli matrices acting in the spin and orbital
subspaces, respectively, while $s_0$ and $\sigma_0$ denote the corresponding
identity matrices. The first five terms constitute the standard four-band
topological-insulator model~\cite{Zhang2009,Liu2010}, whereas the last term
describes the $d$-wave altermagnetic order~\cite{Fu2026}.
In the strong-TI regime considered here, $M_0<0$ and $t_n>0$. Although the
conserved momenta are small near the $\Gamma$ point, the quadratic term along
the surface-normal direction must be retained because it determines the
spatial decay of the surface states.

\subsection{Surface-State Projection on the y Facets}

To derive the leading-order surface Hamiltonian, we first solve the boundary
problem at $t_J=0$ and $k_x=k_z=\mu=0$ in the half-space $y<0$, with a
boundary at $y=0$. Because translational symmetry is broken along the surface
normal, we replace $k_y$ by $-i\partial_y$ and decompose
Eq.~\eqref{eq:app_continuum_amti} as
\begin{align}
H_0^y=&\left(M_0-t_y\partial_y^2\right)s_0\sigma_z
-i\lambda_y\partial_y s_y\sigma_x,
\label{eq:surface_unperturbed_y}\\
H_p^y=&\lambda_xk_xs_x\sigma_x+\lambda_zk_zs_z\sigma_x
-\mu s_0\sigma_0
\nonumber\\
&+\left(t_xk_x^2+t_zk_z^2\right)s_0\sigma_z
+t_J\left(k_x^2+\partial_y^2\right)s_z\sigma_z.
\label{eq:surface_perturbation_y}
\end{align}
Here $H_0^y$ determines the transverse profile of the surface states, whereas
$H_p^y$ contains the finite in-plane momenta, the chemical potential, and the
altermagnetic term. The low-energy surface Hamiltonian is obtained by
projecting $H_p^y$ onto the zero-mode subspace of $H_0^y$. In particular, we
retain the direct projection of the altermagnetic term to leading order in
$t_J$ and neglect $t_J$-dependent corrections to the transverse wave
functions.

The unperturbed Hamiltonian anticommutes with the chiral operator
$C_y=s_y\sigma_y$. For the parameter convention $t_y,\lambda_y>0$, the
normalizable zero modes in the half-space $y<0$ belong to the $C_y=+1$
subspace. In the oscillatory-decay regime
$|M_0/t_y|>\lambda_y^2/(4t_y^2)$ considered in this work, they take the
form~\cite{liu2011,yan2018,pan2024,chen2025}
\begin{equation}
\psi_i^{+}(\bm r)=\frac{1}{N_0}
\sin(\alpha y)e^{\beta y}e^{ik_xx+ik_zz}\chi_i^{+},
\qquad i=1,2.
\end{equation}
For compactness, we introduce
$|\eta\eta'\rangle_y\equiv
|s_y=\eta\rangle\otimes|\sigma_y=\eta'\rangle$ for $\eta,\eta'=\pm1$.
The normalized spinors are
\begin{align}
\chi_1^{+}&=\frac{1}{\sqrt{2}}
\left(|++\rangle_y+|--\rangle_y\right),\\
\chi_2^{+}&=-\frac{i}{\sqrt{2}}
\left(|++\rangle_y-|--\rangle_y\right).
\end{align}
Substitution into Eq.~\eqref{eq:surface_unperturbed_y} gives
\begin{equation}
\alpha=\sqrt{\left|\frac{M_0}{t_y}\right|
-\frac{\lambda_y^2}{4t_y^2}},
\qquad
\beta=\frac{\lambda_y}{2t_y}.
\end{equation}
The normalization factor $N_0$ is given by
\begin{equation}
|N_0|^2=\frac{\alpha^2}{4\beta(\alpha^2+\beta^2)}.
\end{equation}
Since the two surface states share the same transverse envelope, the matrix
elements of $\partial_y^2$ in this subspace are
\begin{equation}
\left\langle\psi_i^{+}\middle|\partial_y^2\middle|\psi_j^{+}\right\rangle
=-\left(\alpha^2+\beta^2\right)\delta_{ij}
=\frac{M_0}{t_y}\delta_{ij}.
\label{eq:app_d2_matrix_element}
\end{equation}
Here $i,j=1,2$ label the two surface states. Let
$P_y^{+}=\sum_{i=1}^{2}|\psi_i^{+}\rangle\langle\psi_i^{+}|$ be the projector
onto the surface-state subspace. In the ordered basis
$(\psi_1^{+},\psi_2^{+})^T$, let $\widetilde s_i$ denote Pauli matrices acting
within this subspace. The remaining projection identities are
\begin{align}
P_y^{+}s_x\sigma_xP_y^{+}&=-\widetilde s_z,
&P_y^{+}s_z\sigma_xP_y^{+}&=\widetilde s_x,\nonumber\\
P_y^{+}s_z\sigma_zP_y^{+}&=\widetilde s_z,
&P_y^{+}s_0\sigma_zP_y^{+}&=0.
\label{eq:app_y_projection}
\end{align}
Using Eqs.~(\ref{eq:app_d2_matrix_element}) and
(\ref{eq:app_y_projection}), and setting $t_y=t_{\parallel}$, the leading-order
surface Hamiltonian is
\begin{equation}
\begin{aligned}
H_{{\rm eff},+}^{y}
&=P_y^+H_p^yP_y^+\\
&=v_zk_z\widetilde s_x-v_xk_x\widetilde s_z-\mu\\
&\quad+\left(t_Jk_x^2+
\frac{t_JM_0}{t_{\parallel}}\right)\widetilde s_z,
\end{aligned}
\label{eq:app_heff_y_plus}
\end{equation}
where $v_x\equiv\lambda_x$ and $v_z\equiv\lambda_z$. For the opposite half-space $y>0$, normalizability instead selects the $C_y=-1$ subspace. Choosing the surface pseudospin basis so that the nonmagnetic Dirac terms have the same form on the two facets gives
\begin{equation}
\begin{aligned}
H_{{\rm eff},\nu}^{y}={}&v_zk_z\widetilde s_x-v_xk_x\widetilde s_z-\mu\\
&+\nu\left(t_Jk_x^2+
\frac{t_JM_0}{t_{\parallel}}\right)\widetilde s_z,
\end{aligned}
\label{eq:app_heff_y_nu}
\end{equation}
where $\nu=\pm1$ labels the two opposite $y$ facets.
Equation~(\ref{eq:app_heff_y_nu}) explicitly shows that the altermagnetic
projection reverses sign between them.

\subsection{Projection on the z and x Facets}
\label{app:z_projection}

The projection is qualitatively different on a $z$ facet. The corresponding unperturbed surface Hamiltonian anticommutes with $C_z=s_z\sigma_y$, while the operator in the altermagnetic term satisfies
\begin{equation}
\left\{C_z,s_z\sigma_z\right\}=0.
\end{equation}
Because the two low-energy states on a given $z$ facet belong to the same
eigenspace of $C_z$, whereas $s_z\sigma_z$ connects opposite $C_z$ sectors,
its direct projection onto the $z$-surface-state subspace vanishes identically,
\begin{equation}
P_zs_z\sigma_zP_z=0,
\qquad
P_zH_JP_z=0.
\label{eq:app_z_projection}
\end{equation}
Thus, within the projected low-energy surface theory, the $d$-wave
altermagnetic term does not contribute to the $z$-surface Hamiltonian. For
$t_x=t_y$ and $\lambda_x=\lambda_y$, the combined $\mathcal{T}C_{4z}$ symmetry
maps the $y$ facets to the $x$ facets. The latter therefore exhibit an
analogous momentum displacement, with $k_x$ replaced by the corresponding
in-plane momentum $k_y$.

\subsection{Finite Net Momentum on the y Facets}

We now determine the finite net momentum on a $y$ facet. For $\nu=+1$, Eq.~\eqref{eq:app_heff_y_nu} gives
\begin{equation}
E(\bm k)=-\mu\pm
\sqrt{(v_zk_z)^2+
\left(v_xk_x-t_Jk_x^2-\frac{t_JM_0}{t_{\parallel}}\right)^2}.
\label{eq:app_y_dispersion}
\end{equation}
At $k_z=0$, selecting the two low-energy Fermi points connected by Andreev reflection yields
\begin{equation}
\begin{aligned}
\delta k_x(\mu)=\frac{1}{2t_J}\bigg[
2v_x
&-\sqrt{v_x^2-4t_J\left(\frac{t_JM_0}{t_{\parallel}}+\mu\right)}\\
&-\sqrt{v_x^2-4t_J\left(\frac{t_JM_0}{t_{\parallel}}-\mu\right)}
\bigg].
\end{aligned}
\label{eq:deltak}
\end{equation}
The high-momentum roots of the quadratic surface dispersion lie outside the low-energy Dirac sector and are not included. On the opposite facet, the corresponding net momentum is $-\delta k_x$. At $\mu=0$, Eq.~\eqref{eq:deltak} reduces to
\begin{equation}
\delta k_x^{(0)}\equiv\delta k_x(0)=
\frac{v_x-\sqrt{v_x^2-4t_J^2M_0/t_{\parallel}}}{t_J}.
\label{eq:deltak0}
\end{equation}
For weak altermagnetic order,
\begin{equation}
\delta k_x^{(0)}\simeq
\frac{2M_0t_J}{v_xt_{\parallel}}.
\end{equation}
To make explicit the weak chemical-potential dependence relevant to the
zero-transverse-momentum Andreev channel, we expand Eq.~\eqref{eq:deltak}
around $\mu=0$. Defining
\begin{equation}
A=v_x^2-\frac{4t_J^2M_0}{t_{\parallel}},
\end{equation}
we obtain
\begin{align}
\delta k_x(\mu)
\simeq&\ \delta k_x^{(0)}
\left[
1+\frac{t_{\parallel}(v_x+\sqrt{A})}
{2A^{3/2}M_0}\mu^2
\right] \nonumber\\
\simeq&\ \delta k_x^{(0)}
\left[
1+\frac{t_{\parallel}}
{v_x^2M_0}\mu^2
\right],
\label{eq:app_deltak_expansion}
\end{align}
where the second line uses $A\simeq v_x^2$ only in the correction factor.
The linear dependence on $\mu$ cancels between the two Fermi points, leaving
the quadratic chemical-potential correction quoted in the main text.

\section{Propagation Phase from the Surface BdG Hamiltonian}
On a $y$ facet, the projected BdG Hamiltonian takes the form
\begin{align}
\mathcal{H}_{\nu}^{y}
=&
\left[
v_z k_z\widetilde{s}_x
-v_x(-i\partial_x)\widetilde{s}_z-\mu
\right]\tau_z
\nonumber\\
&
\quad
+\nu\left[
-t_J\partial_x^2+\frac{t_JM_0}{t_{\parallel}}
\right]\widetilde{s}_z\tau_0
\nonumber\\
&+{\rm Re}\Delta_{\rm ind}(x)\tau_x
-{\rm Im}\Delta_{\rm ind}(x)\tau_y,
\label{eq:app_surface_bdg_pdf}
\end{align}
where $\widetilde{s}_i$ and $\tau_i$ ($i=x,y,z$) are Pauli matrices acting in
the surface-pseudospin and Nambu spaces, respectively, and $\tau_0$ is the
identity matrix in Nambu space. The coefficients $v_x\equiv\lambda_x$ and
$v_z\equiv\lambda_z$ are those introduced in Appendix~A. In this projected
surface description, $\Delta_{\rm ind}(x)$ denotes the proximity-induced pair
potential: it vanishes in the AMTI normal segment and has magnitude
$\Delta_{\rm eff}$ beneath the superconducting electrodes, with the same phase
profile as Eq.~\eqref{eq:pairing_profile}. Here $\Delta_{\rm eff}$ is the
induced surface gap, which is distinct from the parent-superconductor order
parameter $\Delta$ entering the full lattice model. The index $\nu=\pm1$
labels the two opposite $y$ facets and accounts for the opposite projection
of the altermagnetic order. The parameters $M_0$, $t_{\parallel}$, $t_J$, and
$\mu$ retain the meanings defined in Sec.~II. On a $z$ facet, the direct
projection of the altermagnetic term vanishes. The corresponding surface BdG
Hamiltonian therefore has the same Dirac--BdG structure as
Eq.~\eqref{eq:app_surface_bdg_pdf}, with $t_J=0$ and with the transverse term
$v_zk_z\widetilde s_x$ replaced by the appropriate $v_yk_y$ term, up to a
rotation of the surface-pseudospin basis. This effective surface model is used
only to derive the low-energy Andreev quantization condition; the full
numerical calculations explicitly retain the superconducting electrodes and
their coupling to the AMTI.

We first set the transverse momentum to zero. The two helical sectors then
decouple. Propagation of the electron and hole amplitudes through the AMTI
segment is described by
\begin{equation}
T_{\nu}(E)=
\begin{pmatrix}
e^{ik_e^{(\nu)}(E)L_x} & 0\\
0 & e^{ik_h^{(\nu)}(E)L_x}
\end{pmatrix}.
\label{eq:app_propagation_matrix_pdf}
\end{equation}
Here $E$ is the quasiparticle energy and $L_x$ is the length of the AMTI
normal segment. The quantities $k_e^{(\nu)}$ and $k_h^{(\nu)}$ are the
longitudinal wave numbers entering the electron and hole propagation
amplitudes in the BdG scattering basis, respectively. In particular,
$k_h^{(\nu)}$ is a hole wave number and should not be identified directly with
the momentum of the electron state whose absence defines the hole. Following
the notation of Ref.~\cite{Pientka2017}, the scattering matrix at the
superconducting interface $j=L,R$ is written as
\begin{equation}
\begin{aligned}
S_j(E)
&=
e^{i\chi_j\tau_z/2}
S_0(E)
e^{-i\chi_j\tau_z/2},
\\
S_0(E)
&=
\begin{pmatrix}
 r_e & r_A\\
 r_A & r_h
\end{pmatrix}.
\end{aligned}
\label{eq:app_interface_matrix_pdf}
\end{equation}
Here $\chi_j$ is the superconducting phase at interface $j$. In the phase
convention of Eq.~\eqref{eq:pairing_profile}, $\chi_L=-\phi$ and $\chi_R=0$,
so that $\phi=\chi_R-\chi_L$. The quantities $r_e$, $r_h$, and $r_A$ are the
electron normal-reflection, hole normal-reflection, and Andreev-reflection
amplitudes, respectively. These amplitudes are constrained by particle-hole
symmetry and unitarity. We
parametrize them as
\begin{equation}
\begin{aligned}
r_{e/h}
&=\pm r_{\nu}
\exp\left[i\eta_A(E)\pm i\varphi_{N,\nu}\right],
\\
r_A
&=\sqrt{1-r_{\nu}^{\,2}}\,
\exp\left[i\eta_A(E)\right],
\\
\eta_A(E)
&=-\alpha(E),
\alpha(E)=\arccos\left(\frac{E}{\Delta_{\rm eff}}\right),
\end{aligned}
\label{eq:app_reflection_amplitudes_pdf}
\end{equation}
where $r_{\nu}$ is the magnitude of the normal-reflection amplitude,
$\varphi_{N,\nu}$ is the corresponding normal-scattering phase, and
$\eta_A(E)$ is the common reflection phase. Here $\Delta_{\rm eff}$ is the
proximity-induced surface gap defined in the main text, and $\alpha(E)$ is
the Andreev-reflection phase for subgap energies $|E|\leq\Delta_{\rm eff}$.
A subgap bound state must reproduce itself after a complete scattering cycle
consisting of propagation through the normal segment and reflection at the two
superconducting interfaces. Its energy is therefore determined by
\begin{equation}
\det\left[
1-S_L(E)T_{\nu}(E)S_R(E)T_{\nu}(E)
\right]=0.
\label{eq:app_secular_pdf}
\end{equation}
Evaluating Eq.~\eqref{eq:app_secular_pdf} gives
\begin{align}
&\cos\Big\{
\left[k_e^{(\nu)}(E)-k_h^{(\nu)}(E)\right]L_x
+2\eta_A(E)
\Big\}
\nonumber\\
&=
r_{\nu}^{\,2}
\cos\Big\{
\left[k_e^{(\nu)}(E)+k_h^{(\nu)}(E)\right]L_x
+2\varphi_{N,\nu}
\Big\}
\nonumber\\
&+
\left(1-r_{\nu}^{\,2}\right)\cos\phi.
\label{eq:app_general_secular_pdf}
\end{align}
For the helical surface channel considered here, normal reflection is suppressed by spin-momentum locking, such that $r_{\nu}=0$. Setting $\eta_A(E)=-\alpha(E)$, Eq.~\eqref{eq:app_general_secular_pdf}
reduces to
\begin{equation}
\cos\Big\{
\left[k_e^{(\nu)}(E)-k_h^{(\nu)}(E)\right]L_x
-2\alpha(E)
\Big\}
=\cos\phi.
\label{eq:app_transparent_secular_pdf}
\end{equation}
Thus, the normal-region contribution to the Andreev quantization is governed
directly by the electron--hole propagation phase
$\left[k_e^{(\nu)}(E)-k_h^{(\nu)}(E)\right]L_x$.

At zero energy, a BdG hole corresponds to the absence of an electron in the
conjugate helical branch. We denote the two low-energy Fermi points connected
by Andreev reflection by $k_{x,+}^{(\nu)}$ and $k_{x,-}^{(\nu)}$; the labels
$+$ and $-$ distinguish the helical branches and do not imply eigenstates of a
fixed physical-spin component. With the BdG scattering convention used above,
\begin{equation}
k_e^{(\nu)}(0)=k_{x,+}^{(\nu)},
\qquad
k_h^{(\nu)}(0)=-k_{x,-}^{(\nu)},
\label{eq:app_eh_momenta_pdf}
\end{equation}
so that
\begin{equation}
\left[
k_e^{(\nu)}(0)-k_h^{(\nu)}(0)
\right]L_x
=
\left[
k_{x,+}^{(\nu)}
+k_{x,-}^{(\nu)}
\right]L_x.
\label{eq:app_selected_momentum_pdf}
\end{equation}
Evaluating this combination with the projected surface dispersion derived in
Appendix~A gives
\begin{equation}
k_{x,+}^{(\nu)}
+k_{x,-}^{(\nu)}
=\nu\delta k_x,
\qquad
\delta\phi\equiv L_x\delta k_x.
\label{eq:app_am_phase_pdf}
\end{equation}
Therefore, at zero energy,
\begin{equation}
\left[
k_e^{(\nu)}(0)-k_h^{(\nu)}(0)
\right]L_x
=\nu\delta\phi.
\label{eq:app_zero_energy_phase_pdf}
\end{equation}
The scattering equation thus selects the electron--hole momentum combination
$k_e-k_h$, while the projected surface dispersion fixes its zero-energy value
to the facet-dependent phase shift $\nu\delta\phi$. Expanding the propagation
phase to first order in energy gives
\begin{equation}
\left[
k_e^{(\nu)}(E)-k_h^{(\nu)}(E)
\right]L_x
\simeq
\nu\delta\phi+\frac{2EL_x}{\bar v_x},
\label{eq:app_propagation_expansion_pdf}
\end{equation}
where
\begin{equation}
\frac{1}{\bar v_x}
=
\frac{1}{2}
\left(
\frac{1}{|v_{F,+}|}
+\frac{1}{|v_{F,-}|}
\right),
\end{equation}
and $v_{F,+}$ and $v_{F,-}$ are the energy slopes
$\partial E/\partial k_x$ at the two Andreev-connected Fermi points.
Substituting Eq.~\eqref{eq:app_propagation_expansion_pdf} into
Eq.~\eqref{eq:app_transparent_secular_pdf} yields the finite-length
quantization condition
\begin{equation}
\arccos\left(\frac{E}{\Delta_{\rm eff}}\right)
-\frac{EL_x}{\bar v_x}
=
-\zeta\frac{\phi+\nu\delta\phi}{2}
+\pi n,
\label{eq:app_quantization_pdf}
\end{equation}
where $\zeta=\pm1$ labels the particle--hole branches and
$n\in\mathbb{Z}$ is the quantization index. The harmonic mean $\bar v_x$
controls the dynamical propagation phase. In the short-junction limit
\[
L_x\ll \xi_{\rm eff}
=\frac{\bar v_x}{\Delta_{\rm eff}},
\]
where $\xi_{\rm eff}$ is the effective coherence length of the proximitized
surface states. The dynamical phase proportional to $EL_x$ can then be
neglected. Continuously following the signed particle-hole branches gives
\begin{align}
E_{\zeta}^{z}(\phi)
&=
\zeta\Delta_{\rm eff}\cos\left(\frac{\phi}{2}\right),
\label{eq:app_abs_z_pdf}\\
E_{\nu,\zeta}^{y}(\phi)
&=
\zeta\Delta_{\rm eff}
\cos\left[
\frac{\phi+\nu\delta\phi}{2}
\right].
\label{eq:app_abs_y_pdf}
\end{align}
The first expression applies to JJ-$z$, where the altermagnetism-induced
propagation phase vanishes within the projected surface theory. The two
opposite $y$ facets carry $\nu=\pm1$ and therefore acquire opposite phase
shifts. Restoring the transverse Dirac terms to leading order about
$k_\perp=0$ yields, with $v_y\equiv\lambda_y$,
\begin{align}
E_{\zeta}^{z}(k_y,\phi)
&=
\zeta
\sqrt{
(v_y k_y)^2
+\Delta_{\rm eff}^{2}
\cos^{2}\left(\frac{\phi}{2}\right)
},
\label{eq:app_abs_z_ky_pdf}\\
E_{\nu,\zeta}^{y}(k_z,\phi)
&=
\zeta
\sqrt{
(v_z k_z)^2
+\Delta_{\rm eff}^{2}
\cos^{2}\left[
\frac{\phi+\nu\delta\phi}{2}
\right]
}.
\label{eq:app_abs_y_kz_pdf}
\end{align}
Here $\delta\phi=\delta k_xL_x$ is the zero-transverse-momentum
propagation phase defined above; momentum-dependent corrections to this phase
are beyond the leading low-energy Dirac expansion retained in
Eqs.~\eqref{eq:app_abs_z_ky_pdf} and \eqref{eq:app_abs_y_kz_pdf}. Equations~\eqref{eq:app_abs_z_pdf}--
\eqref{eq:app_abs_y_kz_pdf} are the low-energy Andreev bound-state (ABS)
expressions used in the main text. The full lattice calculation retains the
finite junction length, interface coupling, velocity mismatch, and
higher-energy surface and bulk states.

The phase-dependent gaps in these expressions define the effective surface
Dirac masses
\begin{align}
m_z(\phi)&=\Delta_{\rm eff}\cos\left(\frac{\phi}{2}\right)\\
m_{y,\nu}(\phi)&=\Delta_{\rm eff}
\cos\left[\frac{\phi+\nu\delta\phi}{2}\right].
\label{eq:app_surface_masses}
\end{align}
Between the two $y$-facet gap closings,
$m_{y,+}(\phi)m_{y,-}(\phi)<0$. For $\phi\neq\pi$ in this interval,
$m_z(\phi)\neq0$, so exactly one of the two $y$-facet masses has a sign
opposite to that of the adjacent $z$-facet mass. The corresponding hinge is
therefore a domain wall of the superconducting Dirac mass and binds a Majorana
zero mode. This establishes the sequence from the altermagnetism-induced
surface momentum displacement to the facet-dependent Andreev phase, the
surface-mass inversion, and ultimately the hinge MZMs discussed in the main
text.

\bibliographystyle{apsrev4-2}
%apsrev4-2.bst 2019-01-14 (MD) hand-edited version of apsrev4-1.bst
%Control: key (0)
%Control: author (72) initials jnrlst
%Control: editor formatted (1) identically to author
%Control: production of article title (-1) disabled
%Control: page (0) single
%Control: year (1) truncated
%Control: production of eprint (0) enabled
%

% \bibliography{sandwich-junction_Aug_2.bib}
\end{document}